\documentclass[aps,prl,floats,floatfix,twocolumn,longbibliography]{revtex4-1}

\usepackage{color}
\usepackage{latexsym}
\usepackage{amsmath} 
\usepackage{amssymb} 
\usepackage{bm}
\usepackage{wasysym}

\usepackage{graphicx}
\usepackage{epstopdf}
\graphicspath{{./}{./Figs/}}

\usepackage[colorlinks,linkcolor=blue,urlcolor=blue,bookmarks=true,pdftitle={}]{hyperref}

\newcommand{\bra}[1]{\left\langle #1 \right|}
\newcommand{\ket}[1]{\left| #1 \right\rangle}
\newcommand{\braket}[1]{\left\langle #1 \right\rangle }

\newcommand{\beq}{\begin{eqnarray}}
\newcommand{\eeq}{\end{eqnarray}} 

\newcommand{\hide}[1]{}
\newcommand{\Eq}[1]{\textcolor{blue}{Eq.\!\!~(\ref{#1})}} 
\newcommand{\Fig}[1]{\textcolor{blue}{Fig.}\!\!~\ref{#1}}
\newcommand{\sect}[1]{{\bf #1.-- }}

\newcommand{\Cn}[1]{\begin{center} #1 \end{center}}

\newcommand{\hrefl}[2]{\href{#2}{(#1)}}
\newcommand{\rmrk}[1]{#1} 

\begin{document}

\title{Quantum signatures in quench from chaos to superradiance}

\author{Sayak Ray$^{1,2}$,  Amichay Vardi$^{1}$, Doron Cohen$^{3}$ } 

\affiliation{
\mbox{Department of Chemistry, Ben-Gurion University of the Negev, Beer-Sheva 84105, Israel} \\
\mbox{Physikalisches Institut, Rheinische Friedrich-Wilhelms-Universit\"at Bonn, Nu\ss allee 12, 53115, Bonn, Germany} \\
\mbox{Department of Physics, Ben-Gurion University of the Negev, Beer-Sheva 84105, Israel}
}

\begin{abstract}
The driven-dissipative Dicke model features normal, superradiant, and lasing steady-states that may be regular or chaotic. We report quantum signatures of chaos in a quench protocol from the lasing states. Within the framework of a classical mean-field perspective, once quenched, the system relaxes either to the normal or to the superradiant state. Quench-from-chaos, unlike quench from a regular lasing state, exhibits erratic dependence on  control parameters. In the quantum domain this sensitivity implies an effect that is similar to universal conductance fluctuations.  
\end{abstract}

\maketitle


%
The essence of chaos is often presented as a {\em butterfly effect}: a small variation in a control parameter $h$ leads to a drastically different outcome, with seemingly erratic deterministic dependence. For example, a particle is launched into a chaotic cavity and is either transmitted ($Q{=}1$) or reflected ($Q{=}0$). The classical dependence $Q(h)$ looks uncorrelated on a scale that is larger than some exponentially small $\delta h_c$. Alternatively, one may consider  a {\em coin tossing} experiment that involves a dissipative quench to the binary final outcome due to the proverbial coin-ground interaction.

\begin{figure}
\centering
\includegraphics[clip=true, width=6cm]{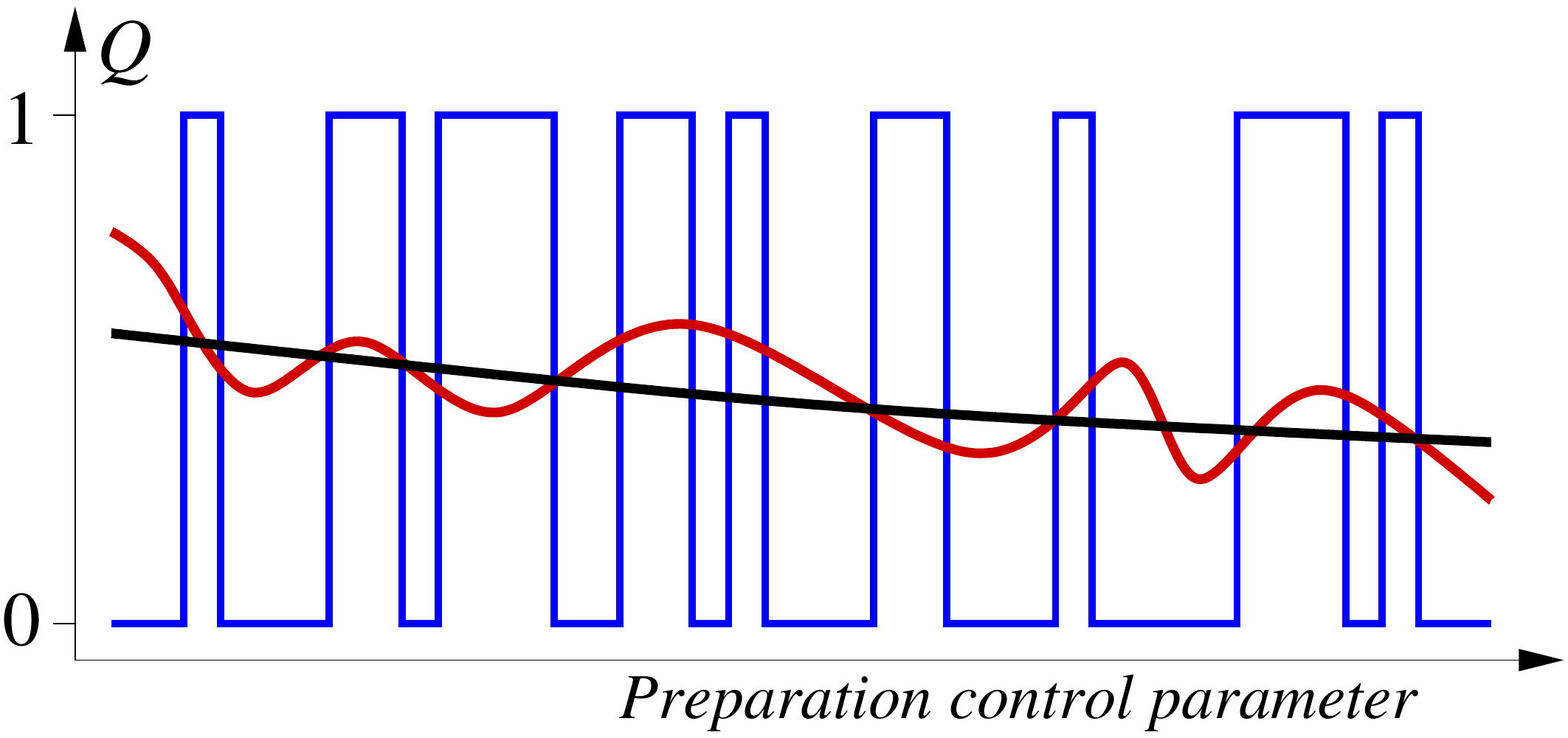}
\caption{{\bf Quantum fluctuations in QFC.}
In the classical (mean-field) limit the outcome of the measurement (blue line) is binary and erratically depends on the parameter that controls the preparation protocol \rmrk{(in our demonstration it is the preparation time $t_{\text{prep}}$)}. In the semiclassical (truncated Wigner) approximation, this erratic dependence is smoothed away (black line). The measured $\braket{Q}$ reflects the relative volume of the basin that leads to the $Q{=}1$ attractor. In the proper quantum treatment the outcome (red line) manifests fluctuations that arise from interference of trajectories. However, any {\em mesoscopic} system eventually relaxes, such that for $t{=}\infty$ the expectation value $\braket{Q}$ reflects a thermal equilibrium that does not depend on the initial preparation.}
\label{fconcept}
\end{figure}

In the present work, we consider a {\em quench from chaos} (QFC) to bistability for atoms in a lasing cavity.  The control parameter $h$ is a pre-quench preparation time $t_{\text{prep}}$, and the post-quench outcome is either a normal state (NS) [$Q{=}0$] or a supperradiant (SR) state [$Q{\ne}0$]. The observable $Q$ is the number of photons in the cavity, namely, $Q=n(t_m)$ where $t_m$ is the time-to-measurement, i.e. the duration of the quench. Within the framework of a classical (Mean Field) perspective, for an appropriate tuning of the atom-field interaction, the dependence of $Q$ on $h$ is erratic, as illustrated in  \Fig{fconcept}. We seek for the signature of this dependence in the quantum regime. 

\rmrk{The simplest quantum version of QFC is a semiclassical phase-space picture}. The wavepacket spreads over the chaotic sea, and therefore the erratic dependence of ${\rm Prob}(Q{=}1)$ on $h$ is smeared away: in the {\em classical} mean-field context this probability is either $0\%$ or $100\%$, while in the {\em semiclassical} truncated Wigner approximation perspective it equals a number $p$ that reflects the relative volume of the basin \rmrk{leading} to the SR state. However, interference between semiclassical trajectories should result in irregular dependence on $h$ in the exact {\em quantum} many-body dynamics, see \Fig{fconcept}.

\rmrk{Fluctuations due to QFC are analogous to universal conductance fluctuations (UCF) \cite{ucf1,ucf2} and chaos-assisted tunneling (CAT) \cite{cat}.} In the UCF context $Q$ is the transmission (conductance) through a chaotic cavity, and $h$ is the magnetic field, while in the CAT context $Q$ is the tunneling rate, and $h$ is the scaled Planck constant. In all those cases (QFC, UCF, CAT) the systematic {\em non-semiclassical} fluctuations in the output signal constitute quantum signature of chaos. \rmrk{However, in QFC we have the extra complication due to dissipation, and one wonders whether any memory of chaos survives after the quench.} The availability of both regular and chaotic lasing steady states in the driven-dissipative Dicke model \cite{Esslinger13, Kirton19, Lesanovsky14, Keeling18, Parkins20, Nori18, Holland10, Barrett17, Thompson12} offers an opportunity to directly contrast the QFC with a quench from a quasi-periodic regular \rmrk{orbit} and show how the $h$ dependence of the quench outcome indicates whether the prepared state was regular or chaotic. 

\sect{Outline}
We first review the regime diagram of the dissipative Dicke model, highlighting NS, SR, as well as regular and chaotic lasing regions. Relaxation towards the NS-SR bistability is then considered as a measurement protocol. In the full QFC scheme, we choose the pre-quench preparation time ($t_{\text{prep}}$) as a control parameter. This QFC scenario is contrasted with the quench \rmrk{from dynamically regular motion}. In particular we aim to clarify the significance of the quench duration ($t_m$).

\sect{The Dicke model}
The model describes $N$ two level atoms (exitation energy $\mathcal{E}$) that interact with a single cavity mode (frequency $\Omega$) \cite{Dicke54, Brandes03}. 
The Hamiltonian involves, respectively, the bosonic field operator $\hat{a}$, 
and the Pauli matrices $\hat{\sigma}_i$, with the common subscripts ${i=x,y,z,\pm}$. 
The couplings $g$ and $\tilde{g}$ denote the strength of the co-rotating and counter-rotating terms of atom-photon interaction. Namely, 
\beq
\hat{H}_{\rm D} &=& \Omega \hat{a}^{\dagger}\hat{a} + \frac{\mathcal{E}}{2} \sum_{r=1}^N\hat{\sigma}_z^r + \frac{g}{\sqrt{N}}\sum_{r=1}^{N} \left(\hat{\sigma}_{+}^r\hat{a}+\hat{\sigma}_{-}^r\hat{a}^{\dagger}\right) \nonumber \\
&+& \frac{\tilde{g}}{\sqrt{N}} \sum_{r=1}^N \left(\hat{\sigma}_{-}^r\hat{a} + \hat{\sigma}_{+}^r\hat{a}^{\dagger}\right)
\label{Dicke_ham}
\eeq
We define the mode oocupation operator ${\hat n=\hat{a}^{\dagger}\hat{a}}$,
and the collective excitation operators $\hat{S}_i^{\ell}=(1/2)\sum_r \hat{\sigma}_i^{r},~(i{=}x,y,z)$ 
that generate a spin algebra with angular momentum $\ell{\leq}N/2$.   

It is well known \cite{Brandes03, Fehske13, Esslinger10, Hemmerich15} that the ground state of the Dicke model undergoes a quantum phase transition from a {\it normal state} (NS) with $\braket{n}=0$ to a pair of {\it superradiant} (SR) states with $\braket{n}\ne0$. 
Moreover, depending on ${(g,\tilde{g})}$, the model exhibits an excited state quantum phase transition \cite{HirschI14, HirschII14}.

\begin{figure}
\centering
\includegraphics[width=\columnwidth]{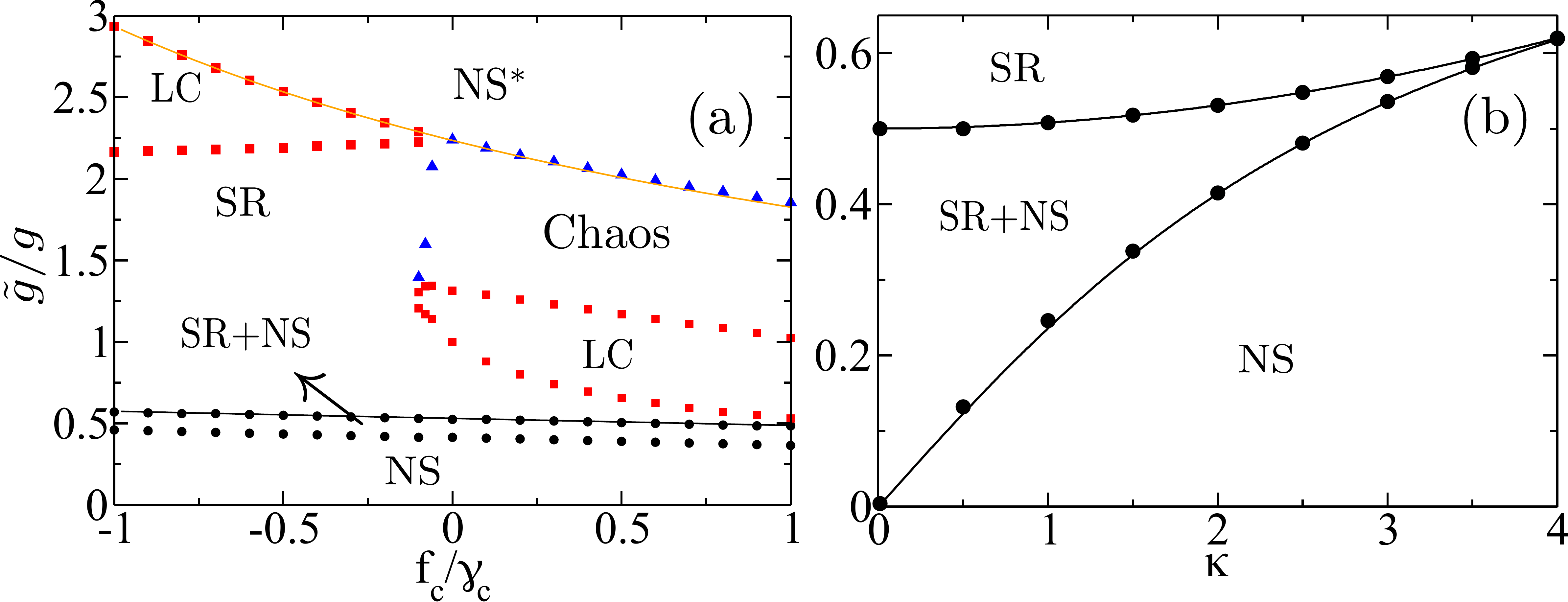}
\caption{{\bf \rmrk{Steady state phase diagram.}}
The vertical axis is the $\tilde{g}/g$ ratio that reflects coherent pumping. 
In panel~(a) the \rmrk{horizontal} axis is the normalized incoherent collective  pumping. 
We assume ${\Omega{=}\mathcal{E}{=}1}$ and $g{=}2$, while $\kappa{=}2$ and $\gamma_c{=}0.5$.
The label NS$^*$ indicates a stable all-atom-excited state. 
The labels LC and Chaos indicate a regular limit cycle  
and a chaotic lasing state, respectively.    
With vanishing dissipation, bistability appears for ${\tilde{g}/g \leq 0.5}$, 
and the energy landscape has 3 attractors (NS and two SR fixed points), 
while with finite dissipation this range is shifted.  
\rmrk{Panel~(b) shows} the dependence of the bistability region on $\kappa$, 
for $g{=}2$, while $f_c{=}\gamma_c{=}0$. 
The symbols are based on numerical analysis,
while the lines are based on stability analysis (see \rmrk{SM}).  
}
\hide{
Panels~(a) and~(b) assume respectively local and collective dissipation mechanism. The vertical axis is the $\tilde{g}/g$ ratio that reflects coherent pumping, while the vertical axis is the normalized incoherent pumping. In both panels we assume ${\Omega=\mathcal{E}=1}$ and ${g=2}$.  
In the limit of zero dissipation, bistability appears for ${\tilde{g}/g \leq 0.5}$, and the energy landscape has 3 attractors (NS and two SR fixed points), while for finite dissipation this range is shifted.       
The symbols are based on numerical analysis,
while the lines are based on stability analysis.  
The label NS$^*$ indicates a stable all-atom-excited state.
The dissipation parameters are $\kappa{=}1$ and $\gamma{=}0.05$ in~(a),
and $\kappa{=}2$ and $\gamma_c{=}0.5$ in~(b).
}
\label{PD_collective}
\end{figure}

\begin{figure}
\centering
\includegraphics[width=\columnwidth]{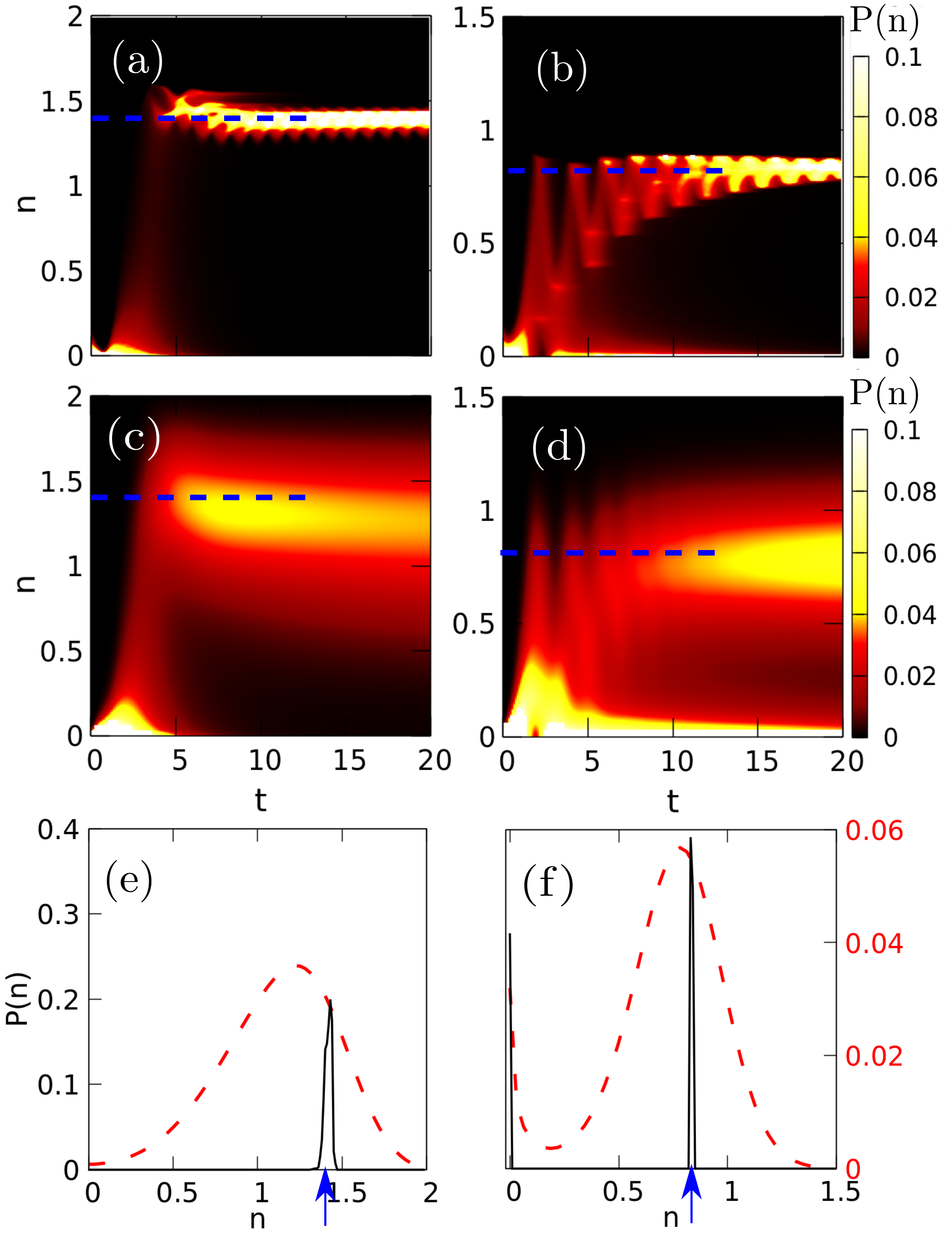}
\caption{{\bf Relaxation towards NS/SR attractors.} 
We start with all the atoms in the ground state, while $n\sim 0$. 
In the left panels $\tilde{g}/g=0.75$, and the relaxation is towards SR. 
In the right panels $\tilde{g}/g=0.48$, and the relaxation is towards NS-SR bistability. 
The other parameters are $g{=}2$, and $\kappa{=}2$, and $\gamma_c{=}0.5$ and $f_c{=}0.04$. 
In the quantum simulation we have \rmrk{$N{=}16$} atoms (meaning \rmrk{$\ell{=}8$}), 
and use $N_b{=}80$ truncation for the bosonic mode.
The semiclassical results of~(a,b) and the quantum results of~(c,d), are compared in~(e,f).  
The waiting time up to the measurement is ${t=t_m=20}$.    
Solid black line is the semiclassical distribution, 
while dashed red line is the quantum distribution.
The classical SR fixed points are marked by horizontal dashed lines in (a-d) 
and by arrowheads in (e-f). Note the $n{=}0$ peak at~(f). }
\label{QCC}
\end{figure}

\sect{Dissipative dynamics} 
Several loss and incoherent processes are associated with the Dicke system \cite{Kirton19, Lesanovsky14, Keeling18, Parkins20, Nori18, Holland10, Barrett17, Thompson12}. The corresponding dissipative dynamics can be studied within the framework of a Lindblad master equation,
\beq
\dot{\rho} &=& -i\left[\hat{H}_\mathrm{D},\rho\right] + \kappa \mathcal{L}[\hat{a}] + \sum_{r=1}^{N} \left(\gamma_{\downarrow}\mathcal{L}[\hat{\sigma}_{-}^r] + \gamma_{\uparrow}\mathcal{L}[\hat{\sigma}_{+}^r]\right) \nonumber \\
&+& \frac{1}{N}\sum_{\ell}^{N/2} \left(\gamma_{\downarrow}^c\mathcal{L}[\hat{S}_{-}^{\ell}] + \gamma_{\uparrow}^c\mathcal{L}[\hat{S}_{+}^{\ell}]\right)
\label{LME}
\eeq
where $\mathcal{L}[\hat{\mathcal{O}}] \equiv \hat{\mathcal{O}}\rho\hat{\mathcal{O}}^{\dagger} - \frac{1}{2}\left(\hat{\mathcal{O}}^{\dagger}\hat{\mathcal{O}}\rho + \rho\hat{\mathcal{O}}^{\dagger}\hat{\mathcal{O}}\right)$.
The incoherent dynamics in \Eq{LME} arises 
from the cavity-photon loss $\mathcal{L}[\hat{a}]$ with a rate $\kappa$, 
and from local incoherent decay and pumping transitions 
$\mathcal{L}[\hat{\sigma}_{-}^r]$ and $\mathcal{L}[\hat{\sigma}_{+}^r]$ 
with rates $\gamma_{\downarrow}$ and $\gamma_{\uparrow}$, respectively. 
Apart from the local incoherent processes, 
there are also incoherent collective processes 
$\mathcal{L}[\hat{S}_{-}^{\ell}]$ and $\mathcal{L}[\hat{S}_{+}^{\ell}]$,
with rates $\gamma_{\downarrow}^c$ and $\gamma_{\uparrow}^c$, respectively. 
%
%
%
%
%
Below, we focus on collective incoherent transitions, and neglect local incoherent processes. The collective decay/pumping for the Dicke model is justified when the atoms are concentrated in a spatial region much smaller than the wavelength of the coupled cavity modes \cite{Kirton19}. The total spin $\ell$ then becomes a constant of motion. Per our \rmrk{preparation} we focus on the $\ell=N/2$ multiplet.
The reduced Hamiltonian can be written in terms of the $S_i$ operators. 
For large $N$ the classical approximation is obtained by treating them as classical coordinates. 
We define scaled variables $s := \hat{S}_{-} /N$, 
and $s_{x,y,z} := \hat{S}_{x,y,z} / N$, 
such that $s_x^2 {+} s_y^2 {+} s_z^2 = 1/4$. 
We also scale the bosonic coordinates \rmrk{as} $a := \hat{a}/\sqrt{N}$. 
Consequently, the classical equations of motion are
\beq
\dot{a} &=& -\left(i\Omega + \kappa/2\right)a - i\left(gs + \tilde{g}s^*\right) \nonumber \\
\dot{s} &=& -\left(i\mathcal{E} + f_cs_z\right)s + 2i\left(ga + \tilde{g}a^{*}\right)s_z \nonumber \\
\dot{s}_z &=& f_c|s|^2 - i\left[g(as^* - a^*s) + \tilde{g}(a^*s^* - as)\right] 
\label{eom_collective}
\eeq
where the net incoherent pumping is  
${f_c = \gamma_{\uparrow}^c - \gamma_{\downarrow}^c}$, 
while the total incoherent rate of transition is 
${\gamma_c = \gamma_{\uparrow}^c + \gamma_{\downarrow}^c}$. 
In \Fig{PD_collective} we present phase-diagrams obtained by stability analysis and numerical long-time propagation of \Eq{eom_collective}. 
%
%
The phase-diagram includes NS, SR, as well as regular and chaotic lasing phases. Moreover, there is a bistable NS-SR phase that we are going to utilize for the measurement protocol.

\begin{figure}
\centering
\includegraphics[width=\columnwidth]{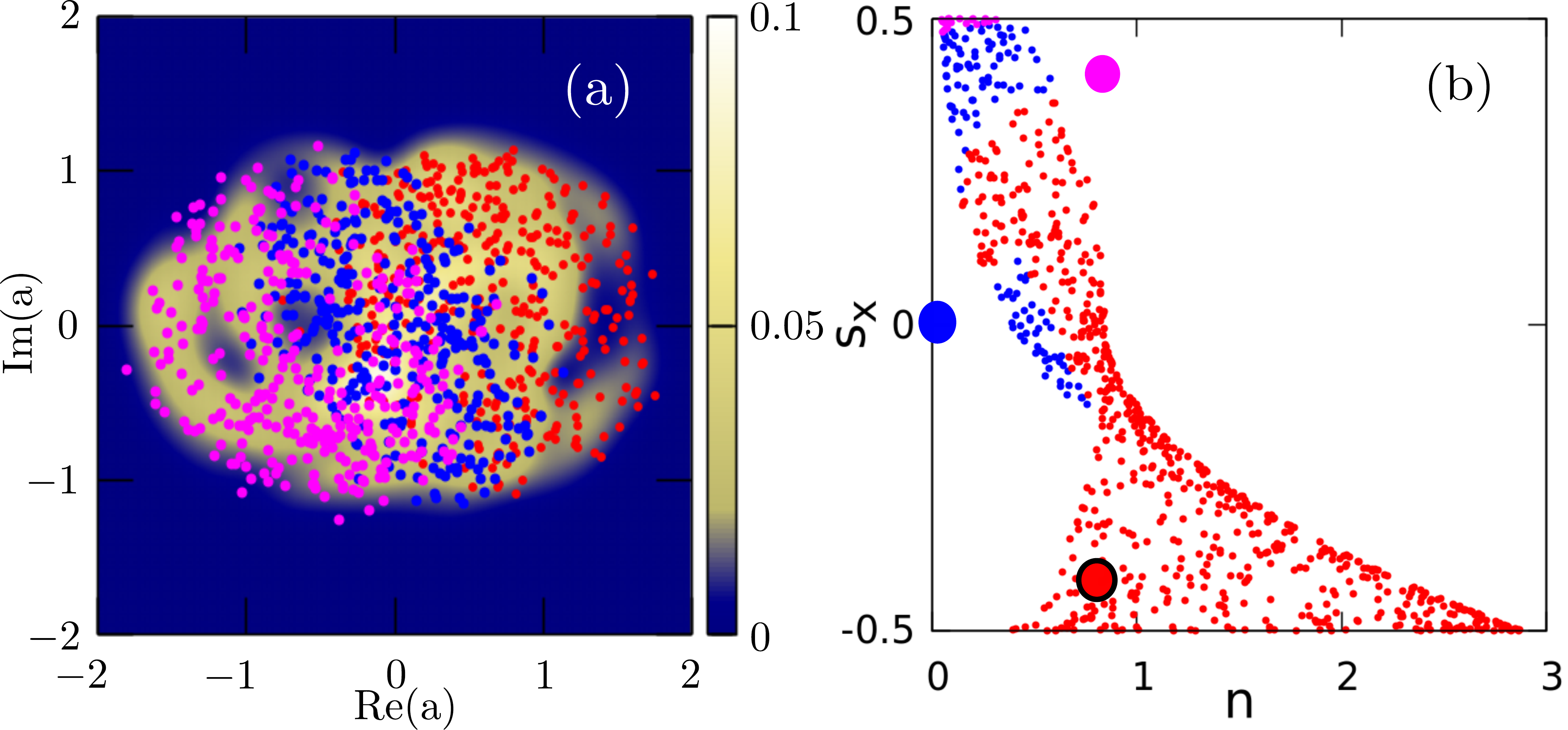}

\caption{{\bf The prepared state.}
The system is prepared in a non-dissipative chaotic state with $g{=}1$ and $\tilde{g}{=}0.48$. 
This is done by launching a coherent state with $s_z{=}s_x{=}1/\sqrt{8}$, and $s_y{=}0$, while ${n {\approx} 0}$,
followed by a long waiting time ${50<t_{\text{prep}}<1000}$.  
In the quantum simulation we have \rmrk{$N{=}16$} atoms (meaning \rmrk{$\ell{=}8$}), 
and use $N_b{=}80$ truncation for the bosonic mode.
Panel~(a) is the quantum Husimi distribution 
of the prepared state in the ${[{\rm Re}(a)-{\rm Im}(a)]}$ plane 
at ${t=t_{\text{prep}}=50}$. 
On top we display the corresponding cloud of classical points. 
The latter are color-coded based on the post-quench outcome: 
blue for those that belong to the NS basin, 
and red/magenta for those of the SR basins. 
Panel~(b) displays the associated $s_z{=}0$ Poincare section (the $s_y,a>0$ branch) 
projected on the $(n-s_x)$ plane,  
with added blue/red/magenta circles that indicate the attractors. 
For the quench we assumed $g{=}2$, but kept the same $\tilde{g}/g$,  
with dissipation parameters $\kappa{=}2$ and $\gamma_c{=}0.5$, 
and with incoherent pumping $f_c{=}0.04$.
}
\label{Husimi}
\end{figure}

\begin{figure}
\centering
\includegraphics[width=\columnwidth]{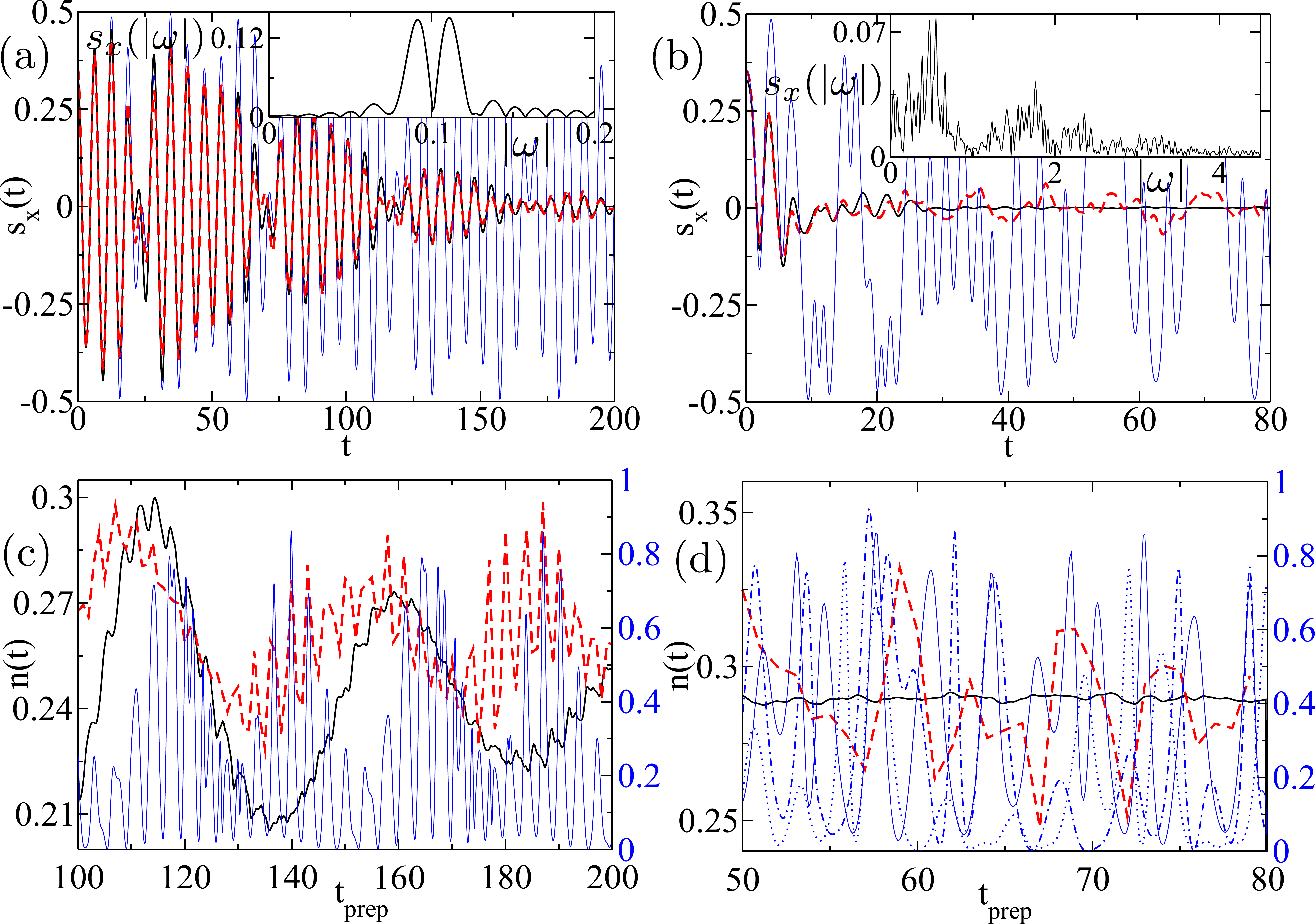}
\caption{{\bf \rmrk{QFC contrasted with non-chaotic dependence.}}
The outcome of a quench versus the control parameter $t_{\text{prep}}$. 
\rmrk{The prepapration assumes disspation-free dynamics.  
Left panels are for a quench from a $g{=}0.1$ quasi-regular, while right panels are for a quench from $g{=}1$ chaos.} 
Panels (a,b) display the pre-quench dynamics of $s_x(t)$. 
The inset displays the associated classical power-spectrum 
(function of $|\omega|$). 
%
Panels (c,d) display the dependence of the quench outcome 
on $t_{\text{prep}}$.  
The quench parameters are the same as in \Fig{QCC},
with measurement time $t_m{=}2$.   
Solid blue and black lines are the classical and semiclassical, respectively, while dashed red line is the quantum.
Panel (d) features the fluctuations that were caricatured in \Fig{fconcept}. 
The solid, dotted and dashed-dotted blue lines in (d) are representative trajectories of the semiclassical cloud in \Fig{Husimi}a exhibiting uncorrelated fluctuations. 
}
\label{Quench}
\end{figure}

\sect{The NS-SR Bistability}
An energy landscape ${E(n,s_z)}$ for the cavity can be obtained by minimizing $H_D$ for a given ${(n,s_z)}$ under the constraint $s_x^2 {+} s_y^2 {+} s_z^2 = 1/4$, see \rmrk{SM}. For small $g$ this landscape exhibits a stable NS minimum at $n{=}0$ and $s_z{=}-1/2$ that becomes an attractor for ${\kappa>0}$. For ${(g{+}\tilde{g}) > \sqrt{\Omega\mathcal{E}}}$, the NS becomes  an energetic saddle point rather than a local minimum, but if ${\tilde{g}/g<1{-}[\sqrt{\Omega\mathcal{E}}/g] }$ it maintains dynamical stability and remains an attractor. The transition of the NS to a saddle point is accompanied by the appearance of a pair of broken symmetry $n {\neq} 0$ SR minima. These two SR states remain attractors provided $\kappa$ is not too large. For quantitative details, including a ${(\kappa, \tilde{g}/g)}$ regime diagram, see \rmrk{SM} \rmrk{and} \Fig{PD_collective}b. 

\sect{Relaxation towards bistability}
In \Fig{QCC} we inspect the distribution $P(n)$ of the cavity mode's occupation. 
In the quantum simulation we start with all the atoms in the ground state, while $n\sim 0$. 
In the semiclassical simulation we prepare an initial cloud centred near the south pole of the Bloch sphere $s_z \sim -1/2$, 
with photon number $n\sim 0$, and let the cloud relax. 
We compare the outcome of relaxation towards a SR steady state,  
to the relaxation in the bistable NS+SR phase.
In the latter case $P(n)$ exhibits two distinct peaks,   
that exhibit broadening in the quantum simulation. 
The quantum SR/NS peak ratio is tilted towards the NS with respect to the classical one due to the quantum spilling from the metastable SR state. It is important to realize that this broadening and peak-ratio tilting are not a signature of true quantum interference: similar broadening would have been captured semiclassically, if Langevin noise terms were included \cite{Gardiner}. By contrast, the quantum-interference signature we seek can not be captured by means of stochastic semiclassical simulations.

\begin{figure}
\centering
\includegraphics[width=\columnwidth]{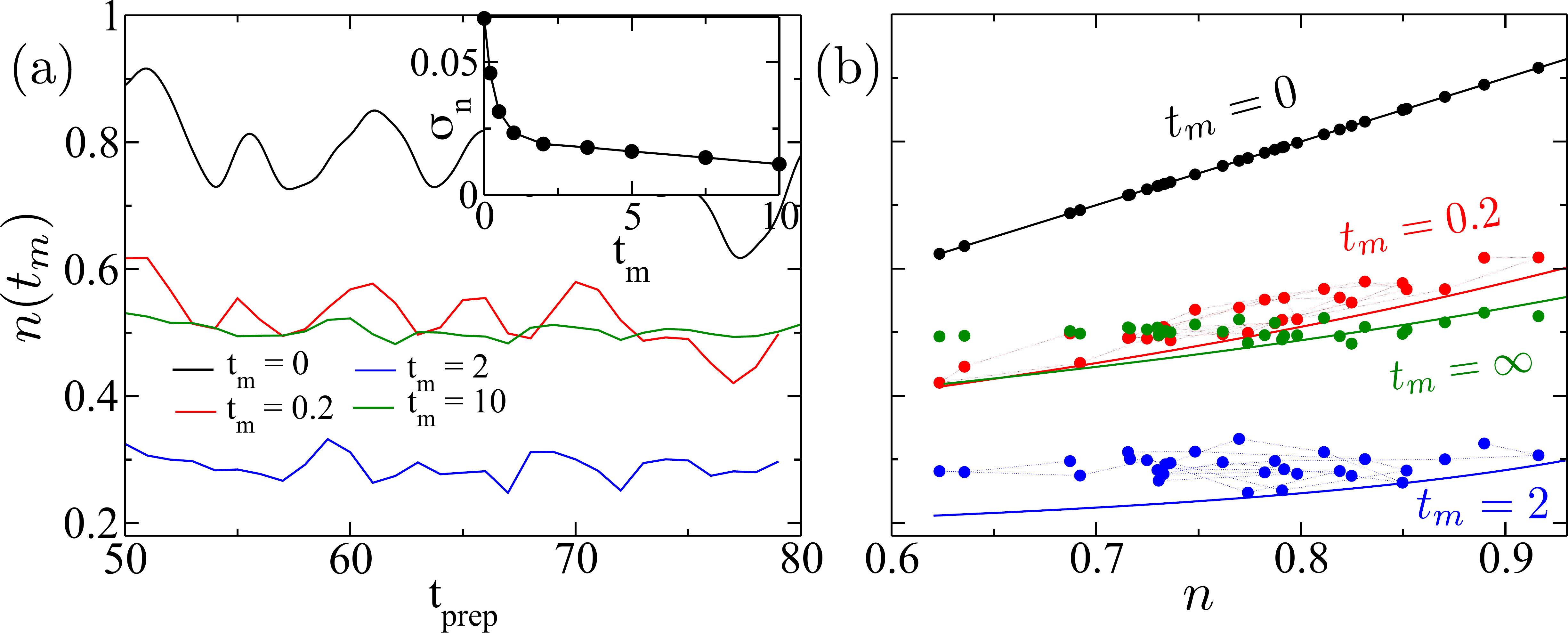}
\caption{{\bf Suppression of quantum fluctuations.}
Panel~(a) displays the dependence of $\langle n(t_m) \rangle$ \rmrk{on} the control parameter $t_{\text{prep}}$ for \rmrk{several} values of~$t_m$. For $t_m{=}0$ it is merely the conventional calculation of $\langle n \rangle$ versus $t$ for the dissipation-free system.
For $t_m{=}\infty$ (in practice $t_m{=}10$) it is formally a measurement of the final equilibrium state.  
The intermediate value $t_m{=}2$, that has been used in \Fig{Quench}, reflects the outcome of a realistic measurement protocol. It exhibits the fluctuations that were caricatured in \Fig{fconcept}. 
\rmrk{The inset shows the variance $\sigma_n$ of those fluctuations  versus $t_m$.}  
The partial correlation between $\langle n(t_m) \rangle$ and $\langle n \rangle$ is inspected in panel~(b), where the data points (symbols) of panel~(a) are connected by thin lines. The thick lines are based on a semiclassical procedure that is explained in the main text. The departure of the data points from the latter is due to relaxation.
}
\label{QFC}
\end{figure}

\sect{Quench from chaos (QFC)}
Having gathered all the necessary ingredients, we turn to discuss the full scenario, including a preparation stage and a quench stage. The purpose of the measurement is to detect chaos in the preparation stage. The quench is to a bistable phase in order to amplify small fluctuations in the prepared state. 

{\em The preparation} of the chaotic state is demonstrated in \Fig{Husimi}. Panel~(a) demonstrates qualitatively the rather good correspondence that we have between the quantum distribution and the semiclassical cloud. The points are color-coded according to which basin they belong: upon quench the blue points will reach the NS fixed point, while the red/magenta points will reach the two SR fixed-points. The phase-space location of the basins is better resolved in the Poincare section of panel~(b).  

The {\em quench} is an abrupt change in the model parameters.
Specifically we force the system to relax towards bistability by setting the parameters ${(g,\tilde{g},\kappa,\gamma_c,f_c)}$ to \rmrk{the values specified for} \Fig{QCC}b. \rmrk{This is followed by a wait time $t_{m}$, during which the system evolves under the dissipative dynamics with the new parameters.}      
\rmrk{At the end of the waiting time, a {\em measurement} of ${Q=\hat{n}(t_m)}$ is preformed. Zero quench time (${t_m{=}0}$) formally means that there is no quench process, 
and accordingly the observable is $Q= \hat{n}(0) = \hat{n}$.} 

For \rmrk{sufficiently} large $t_{m}$, disregarding the quantum/noisy broadening effect, \rmrk{the measured quantity is a sum of a projector on the NS basin, and a projector on the SR basin, weighted by ${n_{\text{NS}}=0}$ and ${n_{\text{SR}}\ne 0}$:}
\beq \label{eQ}
Q = \hat{n}(\infty)
= \sum_{r\in \text{NS}} \ket{r} n_{\text{NS}} \bra{r} + \sum_{r\in \text{SR}} \ket{r}  n_{\text{SR}} \bra{r} 
\ \ \ \ 
\eeq
\Fig{Quench} contrasts the outcome of a QFC with the outcome of a quench from \rmrk{quasi-periodic regular dynamics}.  
The time-to-measurement is intermediate (${t_m=2}$).
We clearly see that chaos is reflected in the outcome of the QFC scenario, in accordance with the discussion of \Fig{fconcept}.
\rmrk{In contrast, the flutuation due to quench from a regular state, 
are non-erratic and merely reflect the spectral context of the quai-regular dynamics.}

\sect{Memory loss}
In a mesoscopic device the information is eventually blurred due to noisy hopping between the fixed points. The outcome of the measurement is presented in \Fig{QFC}a for several choices of $t_{m}$. 
We observe memory loss \rmrk{gradually with increasing $t_{m}$}. For short $t_{m}$ the systematic variation of $Q$ as a function of $t_{\text{prep}}$ is apparent. Furthermore, due to our choice of observable, the outcome is partially correlated with the $t_{m}{=}0$ measurement of $\braket{n}$.  
This is demonstrated in  \Fig{QFC}b. 
We would like to provide a semiclassical procedure for the analysis of this correlation.

In the semiclassical simulation, the ergodized cloud does not show any fluctuations, and therefore, the post-quench dynamics does not depend on the preparation time. However, we can {\em mimic} the quantum fluctuations by giving each ``point" of the semiclassical cloud 
a weight ${w_j \propto (1+Cn_j) }$, where the proportionality constant is determined such that ${\sum w_j=1}$. 
Using the semiclassical equations of motion we can determine the mapping ${n_j \mapsto n_j(t_m)}$. Then we can calculate 
\beq
\braket{Q} \ = \ \braket{n(t_m)}_{sc} \ = \ \sum_j w_j n_i(t_m)
\eeq 
For each $t_{\text{prep}}$ the parameter $C$ is adjusted such that $\braket{n(0)}_{sc}=\braket{n}_{qm}$.  
Then we can predict the outcome for finite $t_m$. 
The result of this phenomenological theory is incorporated in \Fig{QFC}b. 
The departure of the symbols from the calculated lines (e.g. blue as opposed to red symbols) is the signature that fluctuations over the $Q$ of \Eq{eQ} do not reflect trivially fluctuations of~$n$. 
On the other hand, the memory loss due to noisy hopping between the fixed-points
is reflected by the ``flattening" of the outcome (e.g. green symbols).

\sect{Discussion}
A realistic measurement, unlike an idealized projective measurement, involves a dissipative quench process. In a macroscopic reality a tossed-coin, or a ferromagnetic pointer, will always point ``up" or ``down" at the end of the quench. For a non-violent quench, a relatively large $t_{m}$ is required in order to reach the attractor, allowing differentiation between initially similar states. Thermal and quantum fluctuations can be ignored. But in a mesoscopic context, the time of the quench ($t_m$) should be optimized in order to keep the information about the measured (pre-quench) state (it should be ``large" but not too large). Our emphasis was on QFC, looking for the quantum signature of chaos, and clarifying the physical significance of $t_{m}$. Per our construction the ``large" $t_m$ measurement was strongly correlated with the $t_m{=}0$ measurement, but clearly this is not a general feature. In general the ``basins" of $Q$ are not correlated with a simple observable of the system. Either way, we have demonstrated the manifestation of irregular quantum fluctuations in the outcome, providing signature for chaos in the ``measured" state. 
\rmrk{These fluctuations resemble CAT and UCF. They are completely diminished in the semiclassical picture, and come instead of the classical exponential sensitivity that one would expect if reality were not quantum-mechanical. But unlike UCF and CAT, they are endangered by  memory loss due to relaxation.}


\sect{Acknowledgment}
This research was supported by the Israel Science Foundation (Grant No.283/18). SR~acknowledges a scholarship of the Alexander von Humboldt Foundation, Germany.

\clearpage
\onecolumngrid
\pagestyle{empty}

\renewcommand{\thefigure}{S\arabic{figure}}
\setcounter{figure}{0}
\renewcommand{\theequation}{S-\arabic{equation}}
\setcounter{equation}{0}

\Cn{{\large \bf Quantum signatures in quench from chaos to superradiance}} 

\Cn{{Sayak Ray, Amichay Vardi, and Doron Cohen}} 

\Cn{{\large (Supplementary Material)}}

We clarify how the regime diagram of the model is determined. 
The NS and SR steady state solutions of \Eq{eom_collective} 
correspond to the fixed-points ${\dot{a}=\dot{s}=\dot{s}_z=0}$. 
The model parameter are ${(g,\tilde{g},\kappa, f_c)}$. 
Note that $\gamma_c$ implicitly restricts the range of $f_c$, 
but does not appear explicitly in the equations of motion.
The borders of the NS and SR regions in the phase-diagram 
are based on a straightforward linear stability analysis 
of the fixed points that support them. 
Bistability means that there is a region 
where both the NS and the SR fixed-points are stable.

\section{Linear stability analysis}
\label{stability}

Let us denote such steady state (SS) solution by $a^{\rm SS}$, and $s^{\rm SS}$, and $s_z^{\rm SS}$. We consider fluctuations around the SS, namely, $a=a^{\rm SS}+\delta a$, $s=s^{\rm SS}+\delta s$ and $s_z=s_z^{\rm SS}+\delta s_z$. Having put them in \Eq{eom_collective} followed by expanding upto a linear order in $\delta a$, $\delta s$ and $\delta s_z$, we obtain the following set of equations:
\beq
\delta\dot{a} &=& -\left(i\Omega + \kappa/2\right)\delta a - i(g\delta s + \tilde{g}\delta s^{*}) \nonumber \\
\delta\dot{s} &=& -\left( i\mathcal{E} + f_cs_z^{\rm SS} \right)\delta s + 2i\left(g\delta a + \tilde{g}\delta a^{*}\right)s_z^{\rm SS} + 2i(ga^{\rm SS} + \tilde{g}a^{\rm SS *})\delta s_z - f_c s^{\rm SS}\delta s_z \nonumber \\
\delta \dot{s}_z &=& f_c(s^{\rm SS}\delta s^{*}+s^{\rm SS *}\delta s) - i\left(g\delta a + \tilde{g}\delta a^{*}\right)s^{\rm SS *} + i\left(g\delta a^{*} + \tilde{g}\delta a\right)s^{\rm SS} \nonumber \\
&-& i(g\delta s^{*} - \tilde{g}\delta s)a^{\rm SS} + i(g\delta s - \tilde{g}\delta s^{*})a^{\rm SS *} 
\label{fluc}
\eeq
Schematically this set of equations can be written as ${M\Psi=0}$,
where $\Psi \equiv \left[\delta a, \delta a^*, \delta s, \delta s^{*}, \delta s_z\right]^{\mathrm{T}}$ 
is a column vector, and $M$ is a matrix. 
The eigenvalues are determined from the equation ${\det(M-\lambda)=1}$.
Stability of the SS is ensured if all the eigenvalues have negative real part.

\section{NS stability}

At the NS the photon field is zero ($a^{\rm SS}=0$) and the spin polarization is $s_z^{\rm SS}=\pm 1/2$. Thus the linearized equations in \Eq{fluc} decouple from the equation of $\delta s_z$, and hence, the matrix $M$ takes a simple form,
\beq
M=\begin{pmatrix}
-i\Omega-\kappa/2 & 0 & -ig & -i\tilde{g} \\
0 & i\Omega-\kappa/2 & i\tilde{g} & ig \\
2is_z^{\rm SS}g & 2is_z^{\rm SS}\tilde{g} & -i\mathcal{E}-f_cs_z^{\rm SS} & 0  \\
-2is_z^{\rm SS}\tilde{g} & -2is_z^{\rm SS} g & 0 & i\mathcal{E}-f_cs_z^{\rm SS} 
\end{pmatrix}
\eeq
Below we we define $g_{\pm}=(g\pm \tilde{g})$ and ${q=\tilde{g}/g}$. 
The characteristic eigenvalue equation for vanishing dissipation ($f_c=\gamma_c=0$) is 
\beq
\lambda^4 + \kappa \lambda^3 + \left[2g_{+}g_{-} + \mathcal{E}^2 + \Omega^2 + \frac{\kappa^2}{4}\right]\lambda^2 + (g_{+}g_{-} + \mathcal{E}^2)\kappa \lambda + \left[g^4 + \frac{\mathcal{E}^2\kappa^2}{4} + (\tilde{g}^2-\mathcal{E}\Omega)^2 - 2g^2(\tilde{g}^2+\mathcal{E}\Omega)\right] = 0
\ \ \ \ \ \ 
\eeq
The boundary of the NS region is obtained by setting ${\lambda=0}$ which yields,
\beq
\left[4\left(\frac{\mathcal{E}\Omega}{g^2}\right)q^2 - \left(q^2 + \left(\frac{\mathcal{E}\Omega}{g^2}\right) - 1\right)^2\right] = \frac{\kappa^2 \mathcal{E}^2}{4g^4} 
\label{SR-NS_kappa}
\eeq 
In the limit $\kappa \rightarrow 0$, it reduces to
\beq
q^4 - 2q^2 \left[1 + \left(\frac{\mathcal{E}\Omega}{g^2}\right)\right] 
+ \left[1 - \left(\frac{\mathcal{E}\Omega}{g^2}\right)\right]^2 = 0
\label{SR-NS}
\eeq
Given $g$ we get that NS stability survives for 
\beq
q \ \ < \ \ 1 -  \frac{\sqrt{\mathcal{E}\Omega}}{g} 
\eeq
The non-zero range requirement implies 
that $g$ should exceed the critical value $\sqrt{\mathcal{E} \Omega}$
which is relevant for all our numerics. 

\section{SR stability}

The SR fixed-point is the $a^{\rm SS}\ne 0$ fixed point of \Eq{eom_collective}.  
After some algebra, assuming vanishing dissipation ($f_c=\gamma_c=0$), one obtains the following equation for $s_z$, 
\beq
s_z^{2} + \frac{\Omega \mathcal{E} (g_{+}^{2}+g_{-}^{2})}{2g_{+}^2g_{-}^{2}}s_z + \frac{\mathcal{E}^2(\Omega^2+\kappa^2/4)}{4g_{+}^2g_{-}^2}=0
\label{sz_SR}
\eeq
The two solution of the quadratic equation reads,
\beq
s_z = -\frac{1}{2} \left[\frac{\Omega \mathcal{E}}{2g_{+}^2g_{-}^2} (g_{+}^2+g_{-}^2)\right] 
\pm \frac{1}{2} \sqrt{\left[\frac{\Omega \mathcal{E}}{2g_{+}^2g_{-}^2}(g_{+}^2+g_{-}^2)\right]^2 
- \frac{\mathcal{E}^2}{g_{+}^2g_{-}^2} \left(\Omega^2 + \frac{\kappa^2}{4}\right)}
\label{sz_sol}
\eeq
A physical solution of $s_z$ should be a {\it real} number which becomes imaginary at the SR instability. Therefore, vanishing of the term within square-root yields the boundary of the SR region,
\beq
4\Omega^2 q^2 - (1-q^2)^2 \kappa^2 = 0
\label{SR-NS_kappa1}
\eeq
Using the solution of $s_z$ in \Eq{sz_sol}, the corresponding photon field can be obtained from
\beq  
&& {\rm Re}(a)=\pm \sqrt{\left(\frac{1/4-s_z^2}{s_z^2}\right) \times \frac{\mathcal{E}^2/4}{g_{+}^2+r^2g_{-}^2}} 
\\
&& {\rm Im}(a)=\frac{g_{+}^2\kappa s_z}{[\mathcal{E}(\Omega^2+\kappa^2/4)+2g_{-}^2\Omega s_z]} \times {\rm Re}(a)
\eeq
The other two spin components can be obtained from 
\beq
s_x &=& \frac{2 g_{+}}{\mathcal{E}}  {\rm Re}(a) \, s_z \\
s_y &=& -\frac{2 g_{-}}{\mathcal{E}} {\rm Im}(a) \, s_z 
\eeq
It turns out that one of the solutions of \Eq{sz_SR} satisfies the stability criteria in the SR state. At the boundary of the SR region both solutions lose their stability. 

\begin{figure}[b]
\centering
\includegraphics[width=12cm]{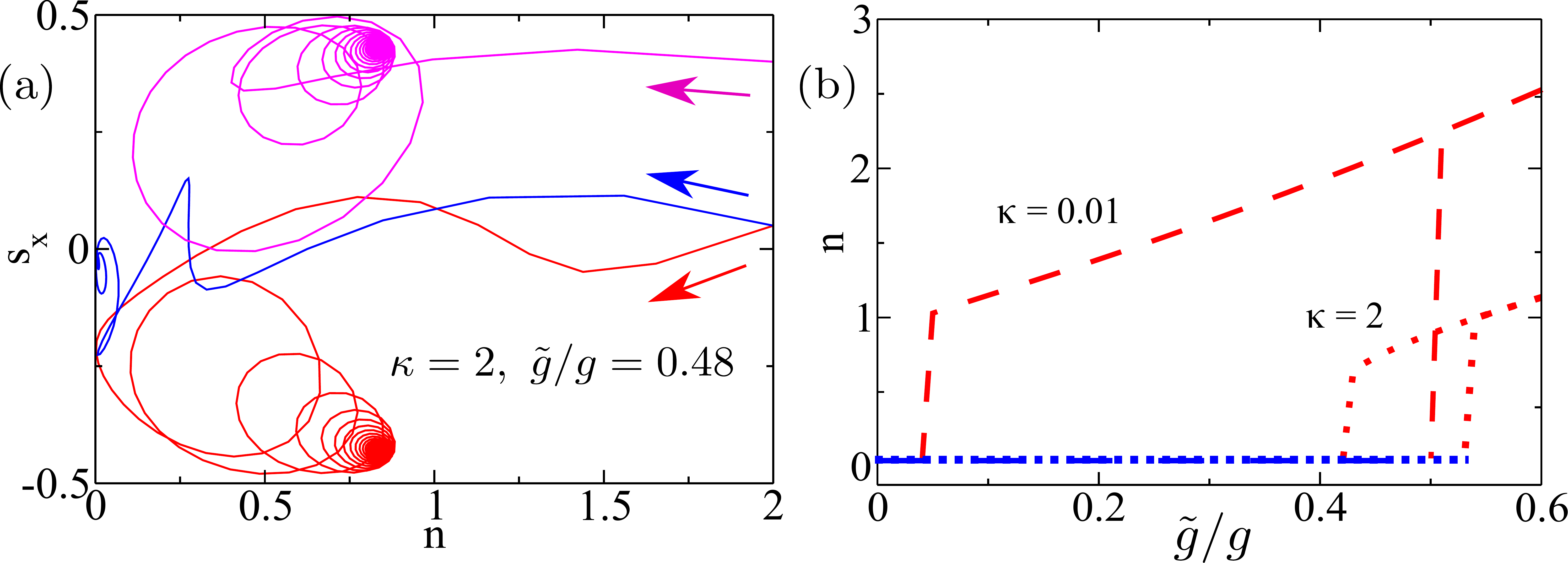}
\caption{{\bf \rmrk{Demonstration of bistability.}}
(a) Demonstration of the dynamics projected on the ${(n,s_x)}$ coordinates. 
Arrows indicate the propagation direction towards the attractors. 
The trajectories that approach the NS and the two SR fixed points  
are colored by blue, magenta and red respectively.  
The interaction is $g{=}2$ and we assume vanishing dissipation ($f_c{=}\gamma_c{=}0$), 
while $\kappa{=}2$.  
(b) The steady state occupation $n$ versus $\tilde{g}/g$ for weak (dashed lines) and strong (dotted lines) cavity losses. 
The SR and the NS steady state values are plotted as red and blue lines, respectively. 
Co-existing blue and red lines indicate bistability region.
}
\label{attractor}
\end{figure}

\section{Bistability}
\label{bistable_kappa}

The common approach to explain the NS-SR symmetry breaking is to look at the energy landscape. 
Using scaled variables as defined in the main text, and ${ a \equiv \sqrt{n} e^{i\varphi} }$, 
the Hamiltonian (divided by $N$) is  
\beq
H(n,\varphi; \vec{s}) \ = \ \Omega n +  \mathcal{E} s_z + 2\sqrt{n} \left[ g_{+} \cos(\varphi) s_x - g_{-} \sin(\varphi) s_y \right]    
\eeq
Given $n$ and the constrain $s_x^2 {+} s_y^2 {+} s_z^2 = 1/4$, 
and assuming that ${\tilde{g}<g}$, the minimum is obtained at $\varphi{=}0$,
and we find
\beq
E(n) \ = \ \text{minimum}\Big\{ H(n,\varphi; s_x,s_y,s_z) \Big\} 
\ = \   \Omega n - \sqrt{ \left(\frac{\mathcal{E}}{2}\right)^2 + g_{+}^2 n} 
\eeq 
We see that the NS fixed-point ($n{=}0$) is no longer the minimum 
if ${ g_{+} >  \sqrt{\Omega\mathcal{E}} }$. 
This is a {\em necessary} condition for bistability. 
The reason for having bistability is that the NS, 
while being a saddle in the energy landscape,
is still a dynamically stable fixed-point, 
that becomes an attractor for finite $\kappa$.       

We plot in \Fig{PD_collective}b the boundaries that we have found for the regions where the NS and the SR fixed points are stable. The two regions {\em overlap}. \rmrk{In the ``SR+NS" overlap region we have bistability, as demonstrated in the simulation of \Fig{attractor}.} 
For vanishing dissipation (${\kappa \rightarrow 0}$) and $g{=}2$ the bistability region is ${0 < \tilde{g}/g < 0.5}$. 
It shrinks with increasing value of $\kappa$, and finally the NS-boundary in \Eq{SR-NS_kappa} touches the SR-boundary in \Eq{SR-NS_kappa1}. Equating \Eq{SR-NS_kappa} and \Eq{SR-NS_kappa1} at the critical $\kappa$ yields
\beq
(q^2-1)\left[4\left(\frac{\mathcal{E}\Omega}{g^2}\right)q^2 - \left(q^2 + \left(\frac{\mathcal{E}\Omega}{g^2}\right) - 1\right)^2\right] - 4\left(\frac{\mathcal{E}\Omega}{g^2}\right)^2q^2 = 0
\eeq
Solving for $q$ for a given $g$ yields $4$ pairs of solution with `$\pm$' counterparts. The lowest positive solution $q^{*}$ and the corresponding $\kappa^{*}$, obtained from \Eq{SR-NS_kappa} togetherly characterizes the maximum of the bistable region. As an example, for $g=2$, as in \Fig{PD_collective}b, $(q^{*},\kappa^{*})$ turns out to be $(0.638,4.3)$.

\ \\ \ \\

\rmrk{\section{Other regimes in the phase diagram}}

\begin{figure}[h!]
\centering
\includegraphics[width=7cm]{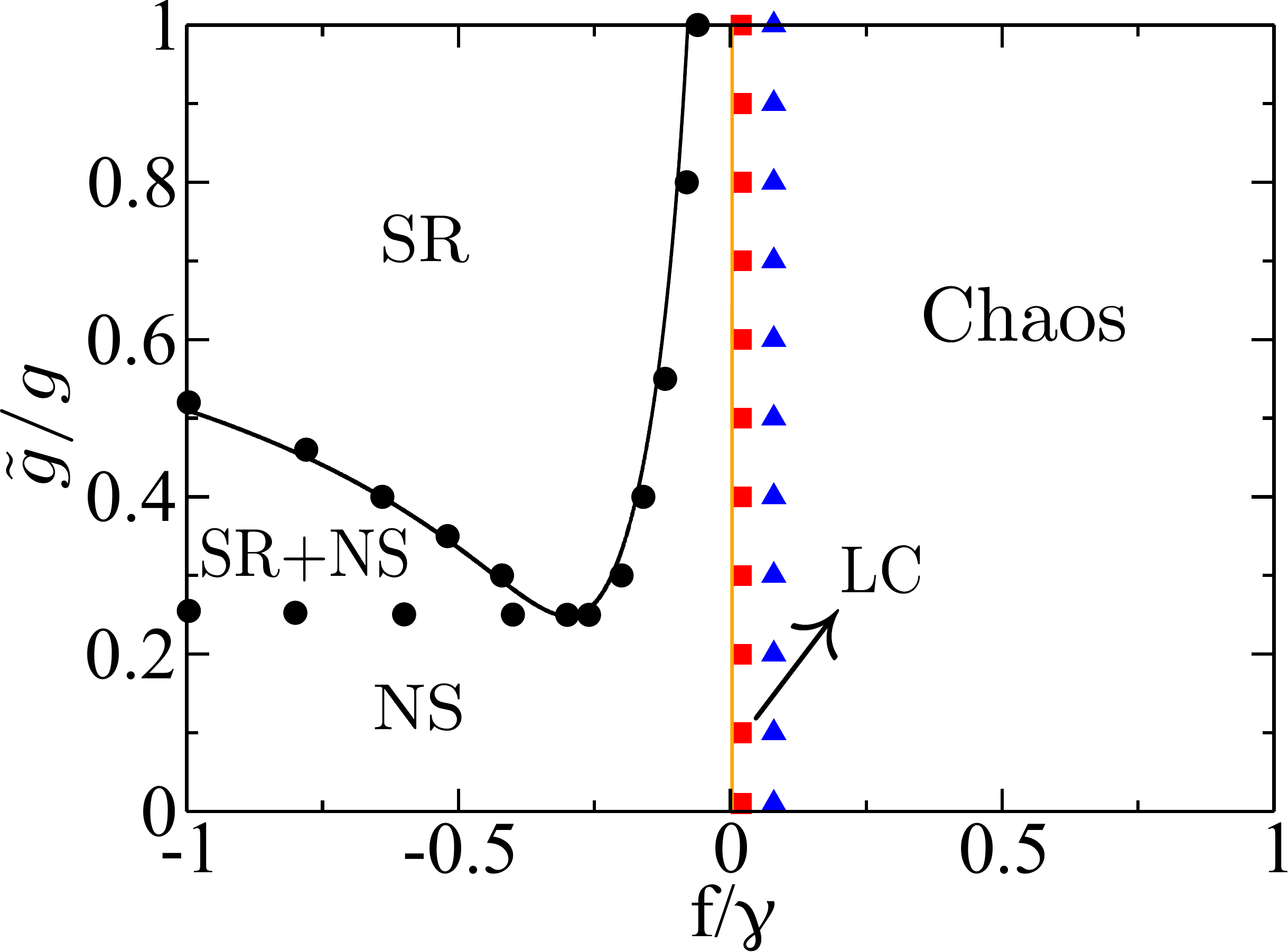}
\caption{{\bf \rmrk{Steady state phase diagram.}}
The vertical axis is the $\tilde{g}/g$ ratio that reflects coherent pumping 
and the horizontal axis is the normalized incoherent local pumping.
We assume ${\Omega{=}\mathcal{E}{=}1}$ and ${g{=}2}$, while $\kappa{=}1$ and $\gamma{=}0.05$. Also here we have NS, SR, LC and chaotic lasing phases. A region with bistability appears as well.}
\label{PD_local}
\end{figure}

Apart from the SR, NS and SR+NS bistable phases, the steady state regime diagram in \Fig{PD_collective}a contains both regular limit cycle (LC) and chaotic lasing state.
Similar observation holds for local dissipation,  as in \Fig{PD_local}, which we display for completeness.

\begin{figure}[h!]
\centering
\includegraphics[width=11cm]{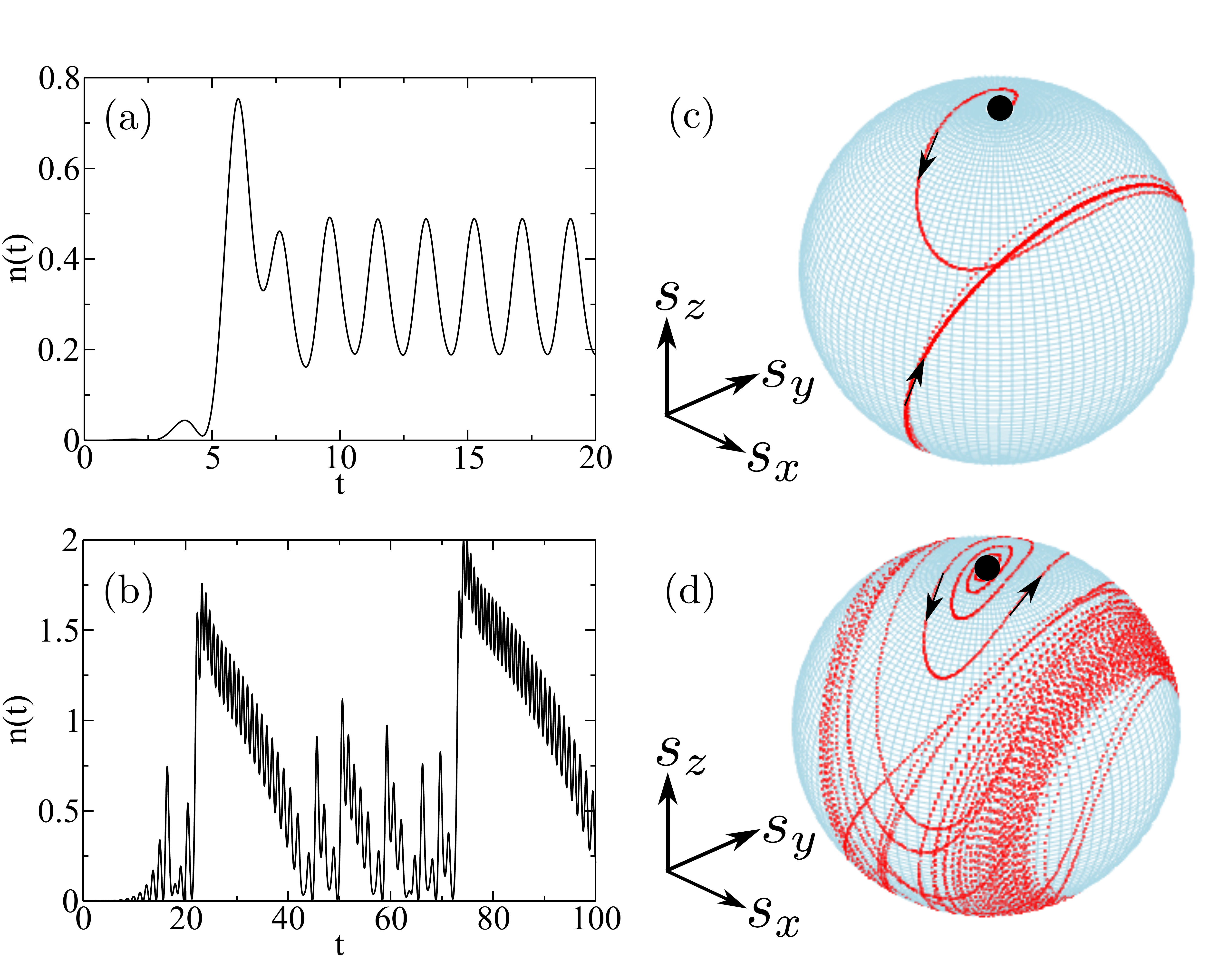}
\caption{{\bf \rmrk{Relaxation towards a lasing state.}}
We plot the photon number $n(t)$ versus time $t$ in the left panel. The corresponding spin dynamics are displayed in the right panel. The initial conditions are $n\approx 0$ and $s_z\approx 1$. The dynamics in the top and bottom panels corresponds to the LC and Chaos phases in \Fig{PD_collective}a with coherent pumping ratio $\tilde{g}/g=1 (1.5)$ and incoherent pumping ratio $f_c/\gamma_c=0.5$.}
\label{lasing_cycle}
\end{figure}

Relaxation towards a lasing steady state is illustrated in \Fig{lasing_cycle}. Depending on the pumping ratios, the system can relax to a chaotic steady state where the output signal from the cavity exhibits irregular oscillations in the photon number. 
Most textbooks focus on the LC regime, where the dissipation is counter balanced by the incoherent pumping. For details one is referred to standard textbooks, e.g. Ref.\cite{Keeling_optics}, and Ref.\cite{Barrett17_supp} for its experimental realization in the context of open Dicke model.

\clearpage

\begin{figure}[h!]
\centering
\includegraphics[width=11cm]{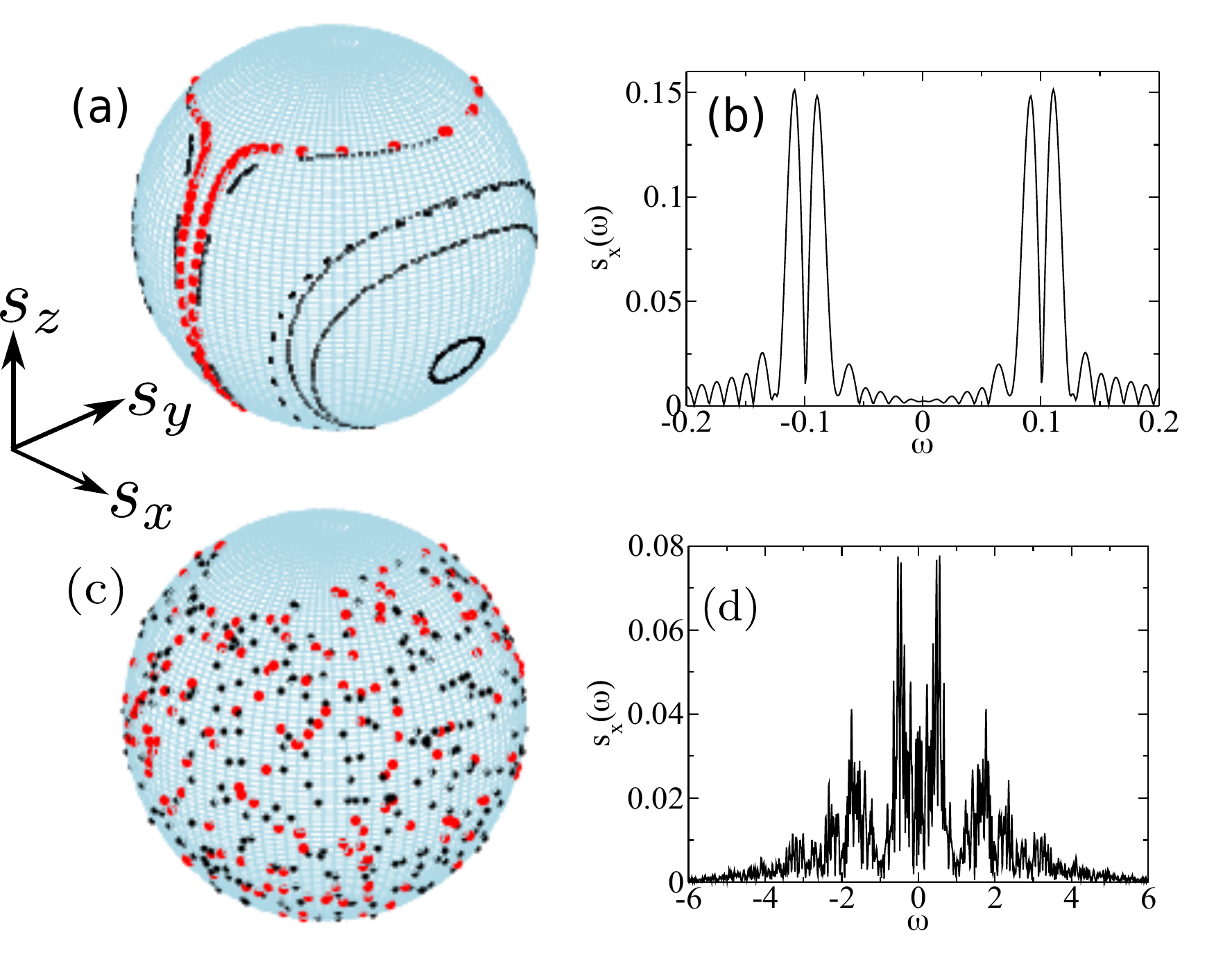}
\caption{{\bf \rmrk{Preparation of pre-quench state.}}
Pre-quench classical spin dynamics are displayed on Bloch sphere with Poincar\'e section at ${\rm Im}(a)=0$ (left panel). The red dotted trajectories correspond to the pre-quench dynamics shown in \Fig{Quench}a,b. Their associated power spectrum obtained from $s_x(t)$ dynamics are shown in the right panel. The top (bottom) panel is prepared for $g=0.1 (1.0)$ with energy $E\approx 0.35 (0.36)$. The other possible islands at the same energy are shown by the black dots.}
\label{dynamics_spin}
\end{figure}

\clearpage
\rmrk{\section{Quench from regular state}}

It is a common practice to characterize {\em chaos} either by the temporal aspect (which justifies the choice of $t_{\text{prep}}$ as a control parameter), or with respect to variation of some other control parameter (magnetic field in the UCF context). 

In \Fig{Quench} we contrasted quench from ``chaos" with quench from ``regular state". We have selected a dissipation-free initial state ($f_c{=}\gamma_c{=}0$). Strictly speaking such states do not reach a steady state. Namely, in the case of chaos, the semiclassical cloud approaches a quasi-ergodic distribution, while regular dynamics typically exhibits damped oscillations due to the broadening of the power spectrum by the nonlinearity.  Still, in the latter case, residual beats are typically observed, and the two dynamical scenarios are readily discernible, as demonstrated in \Fig{Quench}a,b and \Fig{dynamics_spin}. 

For such dissipation-free initial states we were able to simplify the numerical effort enormously, because it was possible to choose $t_{\text{prep}}$ as a control parameter. In the presence of pre-quench dissipation, this choice is not appropriate, because the system always relaxes to a {\em unique} steady state that does not change with time (as implied by the term ``steady state"). Still, in principle, the steady state (regular LC, or chaotic lasing state) can be modified by some other control parameters. Clearly, this opens a wide range of possibilities that can be further studied in the future using the approach that we have proposed in the present work.



\begin{thebibliography}{99}

\bibitem{ucf1}
P.A. Lee and A.D. Stone,
{\em Universal Conductance Fluctuations in Metals},
Phys. Rev. Lett. 55, 1622,
\hrefl{1985}{https://journals.aps.org/prl/abstract/10.1103/PhysRevLett.55.1622}.

\bibitem{ucf2}
R.A. Jalabert,
{\em Mesoscopic transport and quantum chaos},
Scholarpedia, 11(1):30946,
\hrefl{2016}{http://www.scholarpedia.org/article/Mesoscopic_transport_and_quantum_chaos}.

\bibitem{cat}
S. Tomsovic, D. Ullmo, 
{\em Chaos-assisted tunneling},
Phys. Rev. E 50, 145, 
\hrefl{1994}{https://journals.aps.org/pre/abstract/10.1103/PhysRevE.50.145}.

\bibitem{Esslinger13}
H. Ritsch, P. Domokos, F. Brennecke and T. Esslinger,
{\em Cold atoms in cavity-generated dynamical optical potentials},
\rmp {\bf 85}, 553
\hrefl{2013}{https://journals.aps.org/rmp/abstract/10.1103/RevModPhys.85.553}.

\bibitem{Kirton19}
P. Kirton, M. M. Roses, J. Keeling, E. G. Dalla Torre,
{\em Introduction to the Dicke model: from equilibrium to nonequilibrium, and vice versa}, 
Advanced Quantum Technologies, {\bf 2}, 1970013 
\hrefl{2019}{https://onlinelibrary.wiley.com/doi/10.1002/qute.201800043}.

\bibitem{Lesanovsky14}
S. Genway, W. Li, C. Ates, B. P. Lanyon and I. Lesanovsky,
{\em Generalized Dicke Nonequilibrium Dynamics in Trapped Ions},
\prl {\bf 112}, 023603
\hrefl{2014}{https://journals.aps.org/prl/abstract/10.1103/PhysRevLett.112.023603}.

\bibitem{Keeling18}
P. Kirton and J. Keeling,
{\em Superradiant and lasing states in driven-dissipative Dicke models}, 
New J. Phys. {\bf 20}, 015009
\hrefl{2018}{https://iopscience.iop.org/article/10.1088/1367-2630/aaa11d}. 

\bibitem{Parkins20}
K.C. Stitely, A. Giraldo, B. Krauskopf and S. Parkins, 
{\em Nonlinear semiclassical dynamics of the unbalanced, open Dicke model},
Phys. Rev. Research {\bf 2}, 033131 
\hrefl{2020}{https://journals.aps.org/prresearch/abstract/10.1103/PhysRevResearch.2.033131}.

\bibitem{Nori18}
N. Shammah, S. Ahmed, N. Lambert, S. De Liberato and F. Nori,
{\em Open quantum systems with local and collective incoherent processes: Efficient numerical simulations using permutational invariance},
\pra {\bf 98}, 063815
\hrefl{2018}{https://journals.aps.org/pra/abstract/10.1103/PhysRevA.98.063815}.

\bibitem{Holland10}
D. Meiser and M. J. Holland,
{\em Steady-state superradiance with alkaline-earth-metal atoms},
\pra {\bf 81}, 033847
\hrefl{2010}{https://journals.aps.org/pra/abstract/10.1103/PhysRevA.81.033847}.

\bibitem{Barrett17}
Z. Zhiqiang, C. H. Lee, R. Kumar, K. J. Arnold, S. J. Masson, A. S. Parkins and M. D. Barrett,
{\em Nonequilibrium phase transition in a spin-$1$ Dicke model},
Optica {\bf 4}, 424 
\hrefl{2017}{https://www.osapublishing.org/optica/fulltext.cfm?uri=optica-4-4-424&id=362715}.

\bibitem{Thompson12}
J. G. Bohnet, Z. Chen, J. M. Weiner, D. Meiser, M. J. Holland and J. K. Thompson,
{\em A steady-state superradiant laser with less than one intracavity photon}
Nature {\bf 484}, 78 
\hrefl{2012}{https://www.nature.com/articles/nature10920}.

\bibitem{Dicke54}
R. H. Dicke, 
{\em Coherence in Spontaneous Radiation Processes},
Phys. Rev. {\bf 93}, 99 
\hrefl{1954}{https://journals.aps.org/pr/abstract/10.1103/PhysRev.93.99}.

\bibitem{Brandes03} 
C. Emary and T. Brandes, 
{\em Quantum Chaos Triggered by Precursors of a Quantum Phase Transition: The Dicke Model}, \prl {\bf 90}, 044101 
\hrefl{2003}{https://journals.aps.org/prl/abstract/10.1103/PhysRevLett.90.044101}.

\bibitem{Fehske13} 
L. Bakemeier, A. Alvermann and H. Fehske, 
{\em Dynamics of the Dicke model close to the classical limit}, 
\pra {\bf 88}, 043835
\hrefl{2013}{https://journals.aps.org/pra/abstract/10.1103/PhysRevA.88.043835}.

\bibitem{Esslinger10}
K. Baumann, C. Guerlin, F. Brennecke and T. Esslinger, 
{\em Dicke quantum phase transition with a superfluid gas in an optical cavity},
Nature {\bf 464}, 1301 
\hrefl{2010}{https://www.nature.com/articles/nature09009}.

\bibitem{Hemmerich15}
J. Klinder, H. Ke\ss ler, M. Reza Bakhtiari, M. Thorwart and A. Hemmerich,
{\em Observation of a Superradiant Mott Insulator in the Dicke-Hubbard Model},
\prl {\bf 115}, 230403
\hrefl{2015}{https://journals.aps.org/prl/abstract/10.1103/PhysRevLett.115.230403}.

\bibitem{HirschI14}
M. A. Bastarrachea-Magnani, S. Lerma-Hern\'andez and J. G. Hirsch,
{\em Comparative quantum and semiclassical analysis of atom-field systems. I. Density of states and excited-state quantum phase transitions},
\pra {\bf 89}, 032101
\hrefl{2014}{https://journals.aps.org/pra/pdf/10.1103/PhysRevA.89.032101}.

\bibitem{HirschII14}
M. A. Bastarrachea-Magnani, S. Lerma-Hern\'andez, and J. G. Hirsch,
{\em Comparative quantum and semiclassical analysis of atom-field systems. II. Chaos and regularity},
\pra {\bf 89}, 032102
\hrefl{2014}{https://journals.aps.org/pra/abstract/10.1103/PhysRevA.89.032102}.

\bibitem{Gardiner}
C. Gardiner and P. Zoller, {\em Quantum Noise: A Handbook of
Markovian and Non-Markovian Quantum Stochastic Methods
with Applications to Quantum Optics} (Springer Science, 2004).

\end{thebibliography}

\begin{thebibliography}{99}

\bibitem[s1]{Keeling_optics} 
J. Keeling, 
{\em Light-Matter Interactions and Quantum Optics} (CreateSpace Independent Publishing Platform, \hrefl{2012}{https://www.st-andrews.ac.uk/~jmjk/keeling/teaching/quantum-optics.pdf}).

\bibitem[s2]{Barrett17_supp}
Z. Zhiqiang, C. H. Lee, R. Kumar, K. J. Arnold, S. J. Masson, A. S. Parkins and M. D. Barrett,
{\em Nonequilibrium phase transition in a spin-$1$ Dicke model},
Optica {\bf 4}, 424 
\hrefl{2017}{https://www.osapublishing.org/optica/fulltext.cfm?uri=optica-4-4-424&id=362715}.

\end{thebibliography}
\end{document}